\documentclass[aps,prc,twocolumn,superscriptaddress]{revtex4}

\usepackage[dvips]{graphicx,color}
\usepackage{amsmath}
\usepackage{booktabs}

\newcommand{\tr}{\mbox{tr}}

\newcommand{\diff}[2]{\frac{d #1}{d #2}}

\newcommand{\sbrace}[1]{\left( #1 \right)}
\newcommand{\mbrace}[1]{\left\{ #1 \right\}}
\newcommand{\bbrace}[1]{\left[ #1 \right]}

\newcommand{\Slash}[1]{\ooalign{\hfil/\hfil\crcr$#1$}}
\newcommand{\Nucl}[2]{$^{#1}$#2}
\newcommand{\LNucl}[2]{$^{#1}_\Lambda$#2}
\newcommand{\DLNucl}[2]{$^{#1}_{\Lambda\Lambda}$#2}

\newcommand{\PSfig}[2]{\includegraphics[width=#1]{Figs/#2}}
\newcommand{\beq}{\begin{eqnarray}}
\newcommand{\eeq}{\end{eqnarray}}

\newcommand{\btau}{\boldsymbol{\tau}}
\newcommand{\bpi}{\boldsymbol{\pi}}

\newcommand{\VEV}[1]{\langle{#1}\rangle}
\newcommand{\rhoB}{\rho_{\scriptscriptstyle B}}

\newcommand{\Fint}[1]{{\cal D}[#1]}
\newcommand{\chibar}{\bar{\chi}}

\def\nuc#1#2{\relax\ifmmode{}^{#1}{\hbox{#2}}\else${}^{#1}$#2\fi}
\newcommand{\comments}[1]{}
\newcommand{\fSCL}{f_\mathrm{SCL}}
\newcommand{\fpi}{f_\pi}
\newcommand{\fz}{f_\zeta}

\begin{document}


\title{
Lambda hypernuclei and neutron star matter
in a chiral SU(3) relativistic mean field model
with a logarithmic potential
}


\author{K.~Tsubakihara}
\email[]{tsubaki@nucl.sci.hokudai.ac.jp}
\author{H.~Maekawa} 
\author{H.~Matsumiya}
\affiliation{
Department of Physics, Faculty of Science,
Hokkaido University, Sapporo 060-0810, Japan.}
\author{A.~Ohnishi}
\affiliation{
Yukawa Institute for Theoretical Physics,
Kyoto University, Kyoto 606-8502, Japan.}


\date{\today}

\begin{abstract}
We develop a chiral SU(3) symmetric relativistic mean field (RMF) model
with a logarithmic potential of scalar condensates.
Experimental and empirical data of
symmetric nuclear matter saturation properties,
bulk properties of normal nuclei,
and separation energies of single- and double-$\Lambda$ hypernuclei
are well explained.
The nuclear matter equation of state (EOS) is found to be
softened by $\sigma\zeta$ mixing which comes from determinant interaction.
The neutron star matter EOS is further softened by $\Lambda$ hyperons.
\end{abstract}


\maketitle

\section{Introduction}\label{intro}

The equation of state (EOS) of the dense hadronic matter is
one of the keys in nuclear physics
as well as in physics of compact stars
~\cite{Lattimer:1991nc,Shen:1998gq,Lutz:1998uz,
	NS-Hyperons,Glendenning,Schaffner,
	Glendenning:1998zx,Ohnishi:2008ng,Holme:1989ng,Pisarski:1998nh,
	Balberg:1997yw,Sahu:2001um,SchaffnerBielich:2008kb,IOTSY}.
In dense matter, various forms of matter are expected to appear
such as the hyperon admixture~\cite{NS-Hyperons,Glendenning,Schaffner,
	Balberg:1997yw,Sahu:2001um,SchaffnerBielich:2008kb,IOTSY},
meson condensation~\cite{Glendenning:1998zx,Ohnishi:2008ng},
baryon rich quark gluon plasma~\cite{Holme:1989ng},
and color superconductor~\cite{Pisarski:1998nh}.
Among these exotic forms of matter, hyperons are expected to appear
at relatively small densities, $\sim(2-3)\rho_0$.
In addition to the variety of particle species,
partial restoration of chiral symmetry is also expected in nuclear medium,
and it would modify the properties of dense matter significantly.
Therefore, it is desired to respect both hypernuclear physics
information~\cite{Dover:1985ba,Millener:1988hp,Nagara,Bando:1990yi,LambdaData}
and chiral symmetry~\cite{NJL,NJL_PRep,HT01}
in constructing the dense matter EOS.

It is widely believed that hyperons should emerge
such as in the neutron star
core~\cite{NS-Hyperons,Schaffner,
	Balberg:1997yw,Sahu:2001um,SchaffnerBielich:2008kb,IOTSY},
and/or during the black hole formation~\cite{Sumiyoshi:2008kw},
as the baryon density increases in the hadronic matter.
In neutron stars, hyperon admixtures soften the EOS
and reduce the maximum mass of neutron stars.
They also increase the proton fraction,
which may promote faster cooling processes~\cite{Tsuruta:2002ey}.
In black hole formation processes, 
hyperons are abundantly produced due to temperature or density effects,
and shorten the duration time of neutrino emission~\cite{Sumiyoshi:2008kw}.
Hyperon admixtures are governed by the hyperon potentials in nuclear matter,
which may be determined from the hyperon separation energies
from hypernuclei and hyperon production spectra.
From this point of view, we need to adopt a theoretical framework
which can explain both nuclear matter and finite nuclear properties.

A chiral symmetry is another important ingredient in dense matter.
It is a fundamental symmetry of QCD with massless quarks,
and its spontaneous symmetry breaking generates masses of 
constituent quarks and hadrons ~\cite{NJL}.
Hadron properties and EOS would be modified in nuclear matter
due to the partial restoration of the chiral symmetry~\cite{NJL_PRep}.
Thus it is preferable for hadronic many-body theories to possess 
the chiral symmetry, including its spontaneous breaking
and partial restoration at finite densities.

We have recently developed a chiral SU(2) symmetric
relativistic mean field (RMF) model, abbreviated as SCL2 RMF,
with a logarithmic $\sigma$ potential in the form of $-\log\sigma$~\cite{SCL2},
which is derived from 
the strong coupling limit of lattice QCD (SCL-LQCD)~\cite{SCL,SCL-T}.
If we na\"ively include the vector meson in the linear $\sigma$ model
($\phi^4$ theory), the chiral symmetry is found to be restored below 
the normal nuclear density (Lee-Wick vacuum,
or chiral collapse)~\cite{LeeWick_1974,Boguta}.
To avoid this problem, there are several attempts 
some of which result in having instability
at large $\sigma$ values~\cite{Instability} or too stiff EOS~\cite{TooStiff}.
In the SCL2 RMF,
we do not have any instability in the $\sigma$ potential,
and the obtained nuclear matter EOS is found to be reasonably soft.
In addition, the bulk properties of finite nuclei 
(binding energies and charge rms radii) are well described.
Then it is desired to extend the SCL2 to
the SU$_f$(3) version in order to describe hypernuclear systems.
We expect that this extension enables us
to get detailed information of various hypernuclei.

In this paper, we introduce a chiral SU(3) symmetric RMF model,
abbreviated as SCL3 RMF in the later discussion,
as an extension of SCL2 RMF.
%
Some of the model parameters are constrained
by chiral symmetry through the hadron masses and vacuum condensates,
and some of them are determined
by the nuclear matter and finite nuclear properties.
We also determine remaining parameters, meson-$\Lambda$ coupling constants,
by adopting the SU$_f(3)$ symmetric relation for vector couplings
and fitting existing $\Lambda$ hypernuclear data for scalar couplings.
We show that we can reproduce the separation energies of
single $\Lambda$ hypernuclei ($S_\Lambda$)~\cite{LambdaData}
and the $\Lambda\Lambda$ bond energy ($\Delta B_{\Lambda\Lambda}$)
in $^6_{\Lambda\Lambda}$He~\cite{Nagara}
by choosing the coupling constants appropriately.
The EOS of symmetric matter is found to be softened
by the scalar meson with hidden strangeness, $\zeta=\bar{s}s$,
which couples with $\sigma$ through the determinant interaction
representing the effects of U$_A(1)$ anomaly~\cite{KM1970,tHooft1976}.
We also discuss the neutron star matter EOS
and neutron star maximum mass in the present SCL3 RMF.

It is generally preferable to derive the dense matter EOS
from bare baryon-baryon interactions.
Non-relativistic calculations based on the variational method~\cite{FP81,APR}
and the g-matrix~\cite{Nishizaki,Baldo,VPREH00,SPRV06}
have been carried out based on realistic bare baryon-baryon interactions.
These non-relativistic microscopic calculations only with two body forces
do not reproduce the nuclear matter saturation point,
and three-body forces are found to be essential to explain
the saturation property of
symmetric nuclear matter~\cite{FP81,APR,Nishizaki,Baldo,VPREH00,SPRV06}.
In Ref.~\cite{VPREH00,SPRV06},
the modern microscopic \textit{NN}, \textit{YN}, and \textit{YY} interactions
were examined with three-body forces in the framework of Br\"uckner 
Hartree-Fock theory.
The calculated results of the maximum neutron star mass
suggest the importance of the hyperon admixture and the three-body force
in neutron star core.
In the relativistic Br\"uckner Hartree-Fock (RBHF) theory~\cite{RBHF},
empirical nuclear matter saturation is explained quantitatively.
It should be noted that a part of the three-body force effects are 
taken into account in RBHF via Z-type diagrams which come from relativistic
treatments, but RBHF results in Ref.~\cite{RBHF} do not include
the bare three-body forces such as those via baryon resonances.
The EOS in RBHF is approximately reproduced
in RMF with non-linear $\omega$ interaction,
which is introduced to simulate the high density
behavior of the vector potential~\cite{TM1}.
In this work, we follow the latter stand point:
We start from the Lagrangian with several model parameters
and determine these parameters by fitting existing data.
As a result,
the scalar and vector potentials in a symmetric nuclear matter are
found to be consistent
with the RBHF results, then we expect that the results with hyperons
would be also meaningful.

This paper is organized as follows.
In Sec.~\ref{SCL3}, we introduce a chiral SU(3) potential derived from SCL-LQCD
in RMF as an extension from the chiral SU(2) potential.
In Sec.~\ref{Results}, we investigate the properties of
symmetric nuclear matter, normal nuclei, and $\Lambda$ hypernuclei,
and fix the parameters in RMF model
so as to reproduce empirical and experimental data.
Then, we can anticipate neutron star matter EOS
with an RMF model which can explain experimental data.
Finally, we summarize our results and give an outlook.
in Sec. \ref{Summary}.

\section{Chiral SU(3) RMF model}\label{SCL3}

An energy density as a function of the chiral condensate,
abbreviated as a {\em chiral potential} here,
describes the spontaneous chiral symmetry breaking
and its partial restoration through the chiral condensates,
and these chiral condensates determine hadron masses.
In Ref.~\cite{SCL2}, Tsubakihara and Ohnishi proposed to apply
the logarithmic chiral SU(2) potential derived from
the strong coupling limit of lattice QCD (SCL-LQCD)
and developed the SCL2 RMF model.
In Subsec.~\ref{Subsec:SCL2},
we briefly summarize how to derive the chiral potential
in SCL-LQCD~\cite{SCL,SCL2}.
Extension to a chiral SU(3) potential
is described in Subsec.~\ref{Subsec:SCL3},
and a chiral SU(3) symmetric RMF is introduced in Subsec.~\ref{Subsec:SU3RMF}.

\subsection{Chiral SU(2) Potential from SCL-LQCD (SCL2)}\label{Subsec:SCL2}

The lattice QCD action consists of the pure Yang-Mills part
and the fermionic part.
The pure Yang-Mills part is proportional to $1/g^2$, where $g$ is the bare
QCD coupling.
In SCL-LQCD ($g \to \infty$),
we can ignore the pure Yang-Mills action terms,
and only those terms including fermions $S_F$ are kept in the action~\cite{SCL}.
The fermionic action with staggered fermions in the chiral limit
is written in the lattice unit as,
\begin{align}
S_F[\chi,\chibar,U] = &
 \frac12 \sum_{x,\mu}
 	\eta_\mu(x)
	  \left[ \chibar(x)U_\mu(x)\chi(x+\hat{\mu})\right. \nonumber\\
	& \left.-\chibar(x+\hat{\mu})U^\dagger_\mu(x)\chi(x) \right]
 \ ,
\end{align}
where $\eta_\mu(x)=(-1)^{x_0+x_1+\cdots+x_{\mu-1}}$
represents the staggered factor.
After integrating out link variables $U_\mu$
in the leading order of $1/d$ expansion~\cite{SCL},
we obtain the following partition function ${\cal Z}$,
\begin{align}
{\cal Z} =& \int \Fint{\chi,\chibar,U}
\exp\left( -S_F[\chi,\chibar,U] \right)
\nonumber\\
    \simeq& \int\Fint{\chi,\chibar}
	\exp\left[
	\frac12
	\sum_{x,y,\alpha,\beta}
		{\cal M}_{\alpha\beta}(x)V_M(x,y){\cal M}(y)_{\beta\alpha}
	\right] \nonumber\\
         =& \int \Fint{\chi,\chibar,\sigma}
	\exp\left(
		-S_\sigma[\chi,\chibar,\sigma]
	\right)
\label{Eq:Partition}
\ ,\\
S_\sigma =&
	\,\frac12\,
		\sum_{x,y,\alpha,\beta}
		\sigma(x)_{\alpha\beta}V_M(x,y)\sigma(y)_{\beta\alpha}
		\nonumber\\
		&+\sum_{x,y,\alpha,\beta}
			\sigma(y)_{\alpha\beta}
			V_M(y,x)
			{\cal M}(x)_{\beta\alpha}
\ .
\label{auxiliary}
\end{align}
The mesonic composites are defined as
${\cal M}_{\alpha\beta}(x)=\chibar^a_\beta(x)\chi^a_\alpha(x)$,
and the auxiliary fields $\sigma_{\alpha\beta}$ are related to
the expectation values of the mesonic composites,
$\VEV{\sigma_{\alpha\beta}(x)}=-\VEV{{\cal M}_{\alpha\beta}(y)}$.
In these equations, the superscript $a$ denotes color
and the subscripts $\alpha$ and $\beta$ show the flavors of the quark fields.
The lattice mesonic inverse propagator $V_M(x,y)$ is given as
$V_M(x,y)=\sum_\mu\left(\delta_{y,x+\hat{\mu}}+\delta_{y,x-\hat{\mu}}\right)/4N_c$.
From the first to the second line in Eq.~(\ref{Eq:Partition}),
the one-link integral formula,
$\int dU U_{ab}U^\dagger_{cd}=\delta_{ad}\delta_{bc}/N_c$ has been used.

Here, we substitute the auxiliary fields with
the static and uniform scalar $\Sigma_{\alpha\beta}$ and
pseudoscalar $\mathbf{\Pi}_{\alpha\beta}$ matrices
as the mean field ansatz,
\begin{equation}
\sigma_{\alpha\beta}(x)= \Sigma_{\alpha\beta} + i\epsilon(x)\Pi_{\alpha\beta}
\ ,
\end{equation}
where $\varepsilon(x)=(-1)^{x_0+x_1+x_2+x_3}$.
Since fermions are decoupled in each space-time point,
the grassmann integral can be easily evaluated and
the effective free energy is obtained up to a constant as,
\begin{align}
V_\chi(\sigma,\bpi) =& \frac12\VEV{\tr \left[\sigma V_M \sigma\right]}
- N_c\VEV{\log\det(V_M\sigma)}
\nonumber\\
=& \frac{b_\sigma}{2} \mbox{tr}\left[M^\dagger M\right]
- \frac{a_\sigma}{2} \log\det\left[M^\dagger M\right]
\label{Eq:Vchi_SCL_A}
\ ,\\
b_\sigma=&\frac{d}{2N_c}
\ ,\quad
a_\sigma=N_c
\ ,
\end{align}
where $\VEV{\cdots}$ denotes the space-time average,
$d=4$ is the space-time dimension,
and $M$ represents the meson matrix,
$M = \Sigma_{\alpha\beta} + i\Pi_{\alpha\beta}$,
in which the $\varepsilon(x)$ factor is replaced with unity
in $\sigma_{\alpha\beta}$.

While the coefficients $b_\sigma$ and $a_\sigma$
are fixed in the lattice unit in Eq.~(\ref{Eq:Vchi_SCL_A}),
they depend on the lattice spacing
and the scaling factor connecting the meson field and the quark condensate,
which should be chosen for $\sigma$ and $\bpi$ to be in the canonical form.
Furthermore, $n_f$ species of staggered fermions corresponds
to $N_f=4n_f$ flavors, 
and the coefficient modification may not be trivial when we take
$N_f=2$ for SU(2) or $N_f=2+1$ for SU(3).
Thus we stipulate them as parameters to obtain physical meson masses.

In SU(2), meson matrix is given as
$M=(\sigma+i\btau\cdot\bpi)/\sqrt{2}$.
Requiring that the chiral potential has a minimum at $\sigma=f_\pi$
and fitting the pion mass $m_\pi$, one parameter $m_\sigma$ is left
as a free parameter.
Then, the chiral potential is given as,
\begin{align}
V_\chi
  &=  - \frac{a_\sigma}{2}\,\log\left(\det MM^\dagger\right)
      + \frac{b_\sigma}{2}\,\tr(MM^\dagger)
      - c_\sigma\sigma\nonumber\\
  &=  - a_\sigma\,\log(\sigma^2+\bpi^2)
      + \frac{b_\sigma}{2}\,(\sigma^2+\bpi^2)
      - c_\sigma\sigma\nonumber\\
&\simeq
 - 2a_\sigma\,f_\mathrm{SCL}(\frac{\varphi_\sigma}{f_\pi})
      + \frac{1}{2}m_\sigma^2\varphi_\sigma^2
      + \frac{1}{2}m_\pi^2\bpi^2
\label{SU2_EI}
\ ,\\
&f_\mathrm{SCL}(x) = \log(1-x)+x+\frac{x^2}{2}\ ,
\end{align}
where
$\varphi_\sigma=f_\pi-\sigma$,
and the explicit chiral symmetry breaking term $-c_\sigma\sigma$ is introduced.
We have omitted pion self-energy terms and constants
in the third line in Eq.~(\ref{SU2_EI}).
Parameters $a_\sigma$, $b_\sigma$, $c_\sigma$ are given as,
\begin{gather}
a_\sigma={f_\pi^2 \over 4}(m_\sigma^2-m_\pi^2)\ ,\ 
b_\sigma = {1 \over 2}(m_\sigma^2 + m_\pi^2)\ ,\ 
c_\sigma =  f_\pi m_\pi^2\ ,
\label{SU2_EI2}
\end{gather}

With the above logarithmic $\sigma$ potential,
full chiral symmetry restoration is suppressed
because of the repulsive contribution of $V_\chi$ at small $\sigma$.
The present treatment of SCL-LQCD is referred to as the zero temperature
treatment, where $V_\chi$ diverges at $\sigma\to 0$.
In the finite temperature treatment of SCL-LQCD~\cite{SCL-T},
the divergent behavior of $V_\chi$ disappears,
while $V_\chi$ has a finite negative derivative at $\sigma\to 0$.
This finite negative derivative is enough to suppress
full chiral restoration at finite density,
since the nucleon Fermi integral contribution behaves as $\rhoB\sigma^2$
and we always have a minimum at a finite $\sigma$ value.
Therefore, we suppose that the present chiral potential $V_\sigma$
would be a good starting point to investigate cold nuclear matter and nuclei.

\subsection{Chiral SU(3) potential from SCL-LQCD (SCL3)}
\label{Subsec:SCL3}

In order to apply the logarithmic chiral potential to hypernuclear systems,
it is necessary to include mesons with hidden strangeness ($\bar{s}s$)
such as $\zeta$ and $\phi$
in addition to mesons made of $u$ and $d$ quarks
($\sigma$, $\omega$ and $\rho$).
Here, we replace the meson matrix $M$ with that of SU(3),
\begin{gather}
M=
\sbrace{\begin{array}{ccc}
        M_{11}         & a_0^+ + i\pi^+             & \kappa^0 +iK^+ \\
        a_0^- + i\pi^- & M_{22}                     & \kappa^0 +iK^0 \\
        \kappa^- +iK^- & \bar{\kappa}^0 +i\bar{K}^0 & \zeta+i\eta_s
        \end{array}} \ , \\
M_{11} \equiv \frac{1}{\sqrt{2}}\left[
	\sbrace{\sigma+i\eta}
	+\sbrace{a_0^0+i\pi^0}
	\right]
\ ,
\\
M_{22} \equiv \frac{1}{\sqrt{2}}\left[
	\sbrace{\sigma+i\eta}
	-\sbrace{a_0^0+i\pi^0}
	\right]
\ .
\label{MM3d}
\end{gather}

In a similar way to the previous subsection,
the chiral SU(3) potential in SCL-LQCD may be given as,
\begin{align}
V_\chi
  =&- \frac{a'}{2}\,\log\sbrace{\det M'M'^\dagger}
    + \frac{b'}{2}\,\tr\sbrace{MM^\dagger} \nonumber\\
   &- c_\sigma\sigma - c_\zeta\zeta + V_{\mbox{\scriptsize KMT}}
   \ ,
   \label{chiralSU3int}
\end{align}
where the explicit chiral symmetry breaking effects are included
as $c_\sigma\sigma$ and $c_\zeta\zeta$ terms.
Since the strange quark mass is not small compared with $f_\pi$ and $f_\zeta$,
we have taken account of its effects also in the shift of the meson matrix,
$M'=M+\mathrm{diag}(0,0,\delta_s)$.
This shifted meson matrix
plays the role of the constituent quark mass in Eq.~(\ref{auxiliary})
and appears in the logarithmic term of the chiral potential.

The Kobayashi-Maskawa-'t Hooft interaction term~\cite{KM1970,tHooft1976}
is represented in a form of the determinant of meson matrix,
\begin{equation}
V_{\mbox{\scriptsize KMT}}
= -d'\sbrace{\det M + \det M^{\dagger}}
\ .
\label{DetInt}
\end{equation}
This KMT interaction represents the U$_A$(1) anomaly effects.
Without $V_{\mbox{\scriptsize KMT}}$, the above chiral potential is invariant
under $\mathrm{U}_{L}(3)\times \mathrm{U}_{R}(3)$ transformation
in the chiral limit, $\delta_{s} = c_\sigma = c_\zeta = 0$.
In the real world, $\mathrm{U}_A(1)$ symmetry is broken by the anomaly. 
Kobayashi and Maskawa~\cite{KM1970} proposed the above determinant
interaction term,
and this term is derived as the instanton induced quark interaction vertex 
by 't Hooft~\cite{tHooft1976}.

Now we shall decompose the chiral potential $V_\chi$ in Eq.~(\ref{chiralSU3int})
into meson mass terms and interaction terms.
\begin{align}
V_\chi
= & -a'\,\log(\sigma^2\zeta)
	+ \frac{b'}{2}\,(\sigma^2+\zeta^2)-d'\sigma^2\zeta
	-c_\sigma \sigma-c_\zeta \zeta 
\nonumber\\&
	+ \frac12\,\sum_\alpha m_\alpha^2 \phi_\alpha^2
	+ \delta V
\nonumber\\
  = & \frac12\,m_\sigma^2 \varphi_\sigma^2+ \frac12\,m_\zeta^2\varphi_\zeta^2
	+ V_{\sigma\zeta}(\varphi_\sigma,\varphi_\zeta)
\nonumber\\
	+& \frac12\,\sum_\alpha m_\alpha^2 \phi_\alpha^2
	+ \delta V(\varphi_\sigma,\varphi_\zeta,\{\phi_\alpha\})
	+ \mathrm{const.}
\label{chiralSU3int_mass}
\ ,\\
  V_{\sigma\zeta}
  = & - a'\left[
		2\fSCL\left(\frac{\varphi_\sigma}{f_\pi}\right) 
	        +\fSCL\left(\frac{\varphi_\zeta}{f'_\zeta}\right) 
	 \right]
	+ \xi_{\sigma\zeta}\varphi_\sigma\varphi_\zeta
\ ,
\end{align}
where
$\varphi_\sigma=f_\pi-\sigma$ and $\varphi_\zeta=f_\zeta-\zeta$
show the deviation of $\sigma$ and $\zeta$ 
from their vacuum expectation values, respectively,
and $V_{\sigma\zeta}$ denotes the interaction energy density.
In the logarithmic potential, shifted vacuum expectation value of $\zeta$
reads $f'_\zeta = f_\zeta + \delta_s$.
The other meson fields than $\sigma$ and $\zeta$ are shown by $\phi_\alpha$,
and $m_\alpha$ and $\delta V$ represent their masses and interaction terms.
We have ignored the third order term $(\varphi_\sigma)^2\varphi_\zeta$
coming from the determinant interaction.
This term does not change the chiral potential significantly around the vacuum,
but it makes the system unstable at large values of $\sigma$ and $\zeta$.
This is because we do not have polynomial terms such as
$\sigma^4$ and $\zeta^4$,
which stabilizes the chiral potential in the $\phi^4$ theory.
Compared with the case of SU(2),
where all of $\mathrm{tr}(MM^\dagger)$, $\mathrm{det}M$ and 
$\mathrm{det}M^\dagger$ are proportional to the same combination,
$\sigma^2+\bpi^2$,
we have several different terms from $\log(\det M' M'^\dagger)$,
$\mathrm{tr}(MM^\dagger)$ and $\det M+\det M^\dagger$ in SU(3).

In Eq. (\ref{chiralSU3int_mass}), the $\sigma\zeta$ mixing appears
in the quadratic form of the meson fields,
thus we have to diagonalize the mass matrix
to obtain observed $\sigma$ and $\zeta$ meson masses as
\begin{align}
&\frac12
\sbrace{\begin{array}{cc}
        \varphi_\sigma & \varphi_\zeta
        \end{array}}
\sbrace{\begin{array}{cc}
        m_\sigma^2          & \xi_{\sigma\zeta}\\
        \xi_{\sigma\zeta}   & m_\zeta^2
        \end{array}}
\sbrace{\begin{array}{c}
        \varphi_\sigma \\ \varphi_\zeta
        \end{array}} \nonumber\\
&=
\frac12
\sbrace{\begin{array}{cc}
        \varphi_\sigma' & \varphi_\zeta'
        \end{array}}
\sbrace{\begin{array}{cc}
        M_\sigma^2 & 0\\
        0          & M_\zeta^2
        \end{array}}
\sbrace{\begin{array}{c}
        \varphi_\sigma' \\ \varphi_\zeta'
        \end{array}}
\ .
\label{sigzet_diag}
\end{align}

\begin{table*}[bt]
\centering
\caption{Parameters and masses of $a_0$, $\kappa$, $\eta$ and $\eta'$
mesons as functions of $m_\sigma$.
Parameters are determined by fitting
$\pi, K$ and $\zeta$ masses
($m_\pi=138~\mathrm{MeV}$, $m_K=496~\mathrm{MeV}$, $M_\zeta=980~\mathrm{MeV}$),
and vacuum condensate of $\sigma$ and $\zeta$
($f_\pi=92.4~\mathrm{MeV}$, $f_\zeta=94.5~\mathrm{MeV}$).
}
\label{tab:MMT}
\vspace{10pt}
\begin{tabular}{r|rrrrr|rrrrr}
\hline\hline
$m_\sigma$&$m_\zeta$&$a'/\fpi^4$&$b'/\fpi^2$&$d'/\fpi$&$\delta_s/\fpi$&$m_{a_0}$&$m_\kappa$&$M_\eta$&$M_{\eta'}$&$M_\sigma$\\
(MeV)     &(MeV)    &          &          &        &               &(MeV)    &(MeV)     &(MeV)   &(MeV)      &(MeV)     \\
\hline
 630      & 816.7   & 11.06    & 73.07    & 23.81  & 0.4568        & 1108.4  & 1000.8   & 536.0  & 1069.7    & 321.7    \\
 640      & 818.3   & 11.44    & 73.05    & 23.44  & 0.4364        & 1108.3  & 1000.7   & 535.7  & 1062.4    & 344.6    \\
 650      & 819.8   & 11.81    & 73.03    & 23.06  & 0.4175        & 1108.1  & 1000.5   & 535.4  & 1054.9    & 366.4    \\
 660      & 821.4   & 12.20    & 73.01    & 22.67  & 0.4001        & 1108.0  & 1000.3   & 535.0  & 1047.2    & 387.2    \\
 670      & 823.0   & 12.59    & 72.98    & 22.28  & 0.3840        & 1107.8  & 1000.1   & 534.6  & 1039.4    & 407.3    \\
 680      & 824.7   & 12.98    & 72.95    & 21.88  & 0.3690        & 1107.5  &  999.8   & 534.1  & 1031.3    & 426.7    \\
 690      & 826.3   & 13.38    & 72.92    & 21.47  & 0.3550        & 1107.3  &  999.6   & 533.6  & 1023.1    & 445.5    \\
 700      & 828.0   & 13.79    & 72.89    & 21.06  & 0.3421        & 1107.0  &  999.3   & 533.1  & 1014.6    & 463.8    \\
 710      & 829.6   & 14.20    & 72.85    & 20.64  & 0.3300        & 1106.8  &  999.0   & 532.5  & 1006.0    & 481.6    \\
 720      & 831.3   & 14.62    & 72.81    & 20.21  & 0.3187        & 1106.4  &  998.6   & 531.8  &  997.1    & 499.0    \\
 730      & 832.9   & 15.05    & 72.77    & 19.77  & 0.3081        & 1106.1  &  998.2   & 531.1  &  988.0    & 516.0    \\
\hline
Exp.      &         &          &          &        &           &980$\pm$20& 672$\pm$40  & 547.85 & 957.78    & 400-1200 \\
\hline
\hline
\end{tabular}
\end{table*}

Five out of six 
($a'$, $b'$, $c_\sigma$, $c_\zeta$, $d'$ and $\delta_s$)
parameters in this chiral potential
are fixed by fitting observed meson masses
of $m_\pi$, $m_K$ and $M_\zeta$,
and vacuum expectation values of $\sigma$ and $\zeta$ ($f_\pi$ and $f_\zeta$).
Relevant meson masses are related to the parameters $(a', b', d',\delta_s)$ as,
\begin{align}
m_\pi^2 =& b' - \frac{2a'}{\fpi^2}-2d'\fz\ ,\\
m_\sigma^2 =& b' + \frac{2a'}{\fpi^2}-2d'\fz\ ,\\
m_K^2 =& b' - \frac{\sqrt{2}a'}{\fpi\fz'}-\sqrt{2}d'\fpi\ ,\\
m_\zeta^2 =& b' + \frac{a'}{{\fz'}^2}\ ,\\
\xi_{\sigma\zeta} =& - 2d'\fpi\ .
\end{align}
We regard $m_\sigma$ as a model parameter,
and give $a'$ and $b'-2d'\fz$ as in the case of the SCL2 model,
\begin{align}
a'=& \frac{f_\pi^2}{4}\sbrace{m_\sigma^2 - m_\pi^2}
  = a_\sigma
\ ,\\
b'-2d'\fz=& \frac12 \sbrace{m_\sigma^2 + m_\pi^2}
  = b_\sigma
\ .
\end{align}
We assign the observed $\zeta$ as $f_0(980)$,
and the parameters $b', d'$ and $\delta_s$ are determined
to reproduce
$m_\pi=138~\mathrm{MeV}$,
$m_K=496~\mathrm{MeV}$
and $M_\zeta=980~\mathrm{MeV}$.

Coefficients of the linear terms in $\sigma$ and $\zeta$ are determined 
to reproduce the vacuum expectation values,
\begin{align}
c_\sigma =&
\fpi\mbrace{b'- \frac{2a'}{\fpi^2} - 2d'f_\zeta}
= f_\pi m_\pi^2\ ,\\
c_\zeta =& b'\fz- \frac{a'}{f_\zeta'} - f_\pi^2 d'\ .
\end{align}

\begin{figure}
\PSfig{8.5cm}{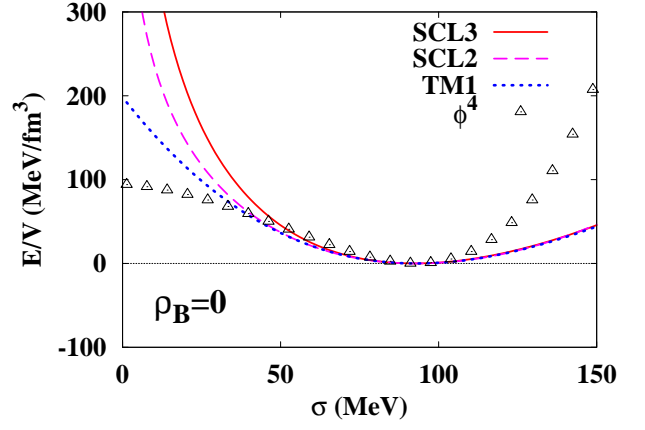}
\caption{
(Color online)
Energy density in vacuum as a function of $\sigma$
in the SCL3 model (solid curve) is compared with those in
the linear $\sigma$ ($\phi^4$, open-triangles),
SCL2 (dashed curve),
and TM1 (dotted curve) models.
}
\label{fig:vac}       
\end{figure}

Once we fix the parameters in the chiral SU(3) potential,
masses of other scalar and pseudoscalar mesons
are determined as shown in Appendix~\ref{AppA}.
Calculated masses of these mesons are tabulated in Table \ref{tab:MMT}.
They are in reasonable agreement with experimental values except
for $\kappa$.

\begin{figure}
\PSfig{8.5cm}{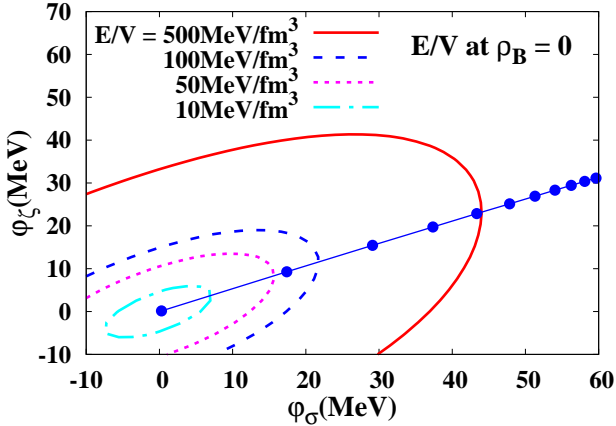}
\caption{(Color online)
Energy surface of chiral effective potential $V_{\sigma\zeta}$
at $\rhoB = 0$ in SCL3.
Points show chiral condensates
($\varphi_\sigma$ and $\varphi_\zeta$)
at finite densities, 
$\rhoB/\rho_0=0, 0.5, \cdots 5$.
}\label{fig:szmix}       
\end{figure}

In Fig.~\ref{fig:vac}, we show the energy density as a function of $\sigma$.
We compare the SCL3 results
with those in SCL2~\cite{SCL2}, TM1~\cite{TM1}
and the linear $\sigma$ ($\phi^4$) models.
We adopt the parameter $m_\sigma=690~\mathrm{MeV}$,
which reproduces the bulk properties of normal nuclei
as explained later,
and optimal $\varphi_\zeta$ value is chosen for each $\sigma$.
When we only consider the quadratic term in $\varphi_\sigma$,
the energy density behaves as $m_\sigma^2\varphi_\sigma^2/2$.
Thus the energy density in SCL3 can be twice larger than the results
in SCL2 and TM1, in which $m_\sigma$ is around 500 MeV,
while the calculated results shows similar values around $\sigma=f_\pi$.
This is because the optimal $\zeta$ value is chosen and reduces the energy
density in SCL3.
In Fig.~\ref{fig:szmix}, we show the energy surface as a function 
of $\varphi_\sigma$ and $\varphi_\zeta$.
The optimal value of $\varphi_\zeta$ is modified from zero to a finite value
by the $\sigma\zeta$ coupling from the KMT interaction,
and the energy density is reduced by this mixing.

\subsection{SCL3 RMF model}\label{Subsec:SU3RMF}

We incorporate the chiral SU(3) potential $V_{\sigma\zeta}$
discussed in the previous subsections into the SU(3) RMF model.
We consider the following SU(3) RMF Lagrangian,
which describes baryons which couple with
$\sigma$ and $\zeta(=\bar{s}s)$ scalar mesons,
and $\omega$, $\rho^0$ (denoted by $R$), and $\phi$ vector mesons,
\begin{align}
\mathcal{L} =
& \sum_i\bar\psi_i \left[ i \Slash{\partial}
	- M^*_i - \gamma_\mu U_i^\mu
      \right] \psi_i\nonumber\\
  +& \frac12\,\partial_\mu\varphi_\sigma\partial^\mu\varphi_\sigma
                    - \frac12\,m_\sigma^2 \varphi_\sigma^2
   + \frac12\,\partial_\mu\varphi_\zeta\partial^\mu\varphi_\zeta 
                    - \frac12\,m_\zeta ^2 \varphi_\zeta ^2
   \nonumber\\
  -& \frac14\omega_{\mu\nu}\omega^{\mu\nu}
   + \frac{m_\omega^2}{2} \omega_\mu\omega^\mu
   - \frac14     R_{\mu\nu}     R^{\mu\nu}
   + \frac{m_\rho^2}{2}    R_\mu  R^\mu
   \nonumber\\
   -& \frac14\phi_{\mu\nu}\phi^{\mu\nu}
   + \frac{m_\phi^2}{2} \phi_\mu\phi^\mu
   - \frac14 F_{\mu\nu} F^{\mu\nu}
   \nonumber\\
   +& \frac{c_\omega}{4}\sbrace{\omega_\nu\omega^\nu}^2
   - V_{\sigma\zeta}\left(\varphi_\sigma, \varphi_\zeta\right)
   \ ,
\label{SU3Lag}
\end{align}
\begin{align}
M^*_i=&M_i - g_{\sigma i}\varphi_\sigma - g_{\zeta i}\varphi_\zeta
\ ,\\
U^\mu_i=&
      g_{\omega i}\,{\omega}^\mu + g_{\rho i}\,R^\mu + \frac{1 + \tau_3}{2}eA^\mu 
\ ,
\label{Eq:VectorPot}
\end{align}
where $V^{\mu\nu} (V=\omega, R, \phi)$ shows
the field tensor of the vector meson $V$.
The $\omega^4$ term in Eq.~(\ref{SU3Lag})
is introduced to simulate the behavior of the vector potential
at high densities in the RBHF theory~\cite{RBHF,TM1}.
The term $V_{\sigma\zeta}(\varphi_\sigma, \varphi_\zeta)$ is
the scalar self-interaction,
and we adopt the chiral SU(3) potential $V_{\sigma\zeta}$
in Eq.~(\ref{chiralSU3int_mass}).

In this Lagrangian, we have several model parameters to be fixed;
$m_\sigma$ in $V_{\sigma\zeta}$,
$c_\omega$ for the $\omega$ self-interaction,
and the meson-baryon coupling constants $g_{mB}$.
We assume that (1) nucleon mass is fully generated by the chiral condensate,
$M_N=g_{\sigma N}f_\pi$,
(2) the vector couplings obey the SU$_f$(3) relation~\cite{Dover:1985ba,M93},
and 
(3) nucleon does not couple with hidden strangeness mesons
($\zeta$ and $\phi$)~\cite{OZI,Isgur:2000ts}.
Suppressed $N\phi$ coupling is understood in the 
Okubo-Zweig-Iizuka (OZI) rule~\cite{OZI},
while the scalar meson-nucleon coupling $g_{\zeta N}$ may violate
the OZI rule~\cite{Isgur:2000ts}.
Thus the assumption of $g_{\zeta N}=0$ may be regarded as
an working hypothesis.

Under these assumptions, we have four meson-baryon coupling constants,
$g_{\omega N}, g_{\rho N}, g_{\sigma\Lambda}, g_{\zeta\Lambda}$,
as model parameters.
We have totally six parameters ($m_\sigma, c_\omega, g_{\omega N}, g_{\rho N},
g_{\sigma\Lambda}, g_{\zeta\Lambda}$).
For the parameters relevant to normal nuclear properties
($m_\sigma, c_\omega, g_{\omega N}, g_{\rho N}$),
first we give $m_\sigma$ and fix $g_{\omega N}$ and $c_\omega$
by fitting the saturation point.
Next from the binding energies of Sn and Pb isotopes,
$m_\sigma$ and $g_{\rho N}$ are obtained.
The separation energies of $\Lambda$ hypernuclei mainly reflects
the core-$\Lambda$ potential depth, $U_\Lambda\sim - 30~\mathrm{MeV}$,
thus the combination
$g_{\sigma\Lambda}\varphi_\sigma(\rho_0)+g_{\zeta\Lambda}\varphi_\zeta(\rho_0)$
is obtained from this fitting procedure.
Finally, the ratio of $g_{\sigma\Lambda}$ and $g_{\zeta\Lambda}$ is 
determined from the $\Lambda\Lambda$ bond energy in the double $\Lambda$
hypernucleus.
All parameters are tabulated in Table \ref{tab:params}.

\begin{table}[bt]
\centering
\caption{Parameter set determined from saturation point of
         symmetric nuclear matter, binding energies and charge rms radii
         of normal nuclei, $s_\Lambda$ of single $\Lambda$ hypernuclei
         and $\Delta B_{\Lambda\Lambda}$ in \DLNucl{6}{He}.
	 Input constants adopted in this paper are also shown.
	 We adopt the saturation point,
	 $(\rho_0,E_0/A)=(0.150~\mathrm{fm}^{-3}, -16.3~\mathrm{MeV})$. 
	}\label{tab:params}
\vspace{10pt}
\begin{tabular}{ccccccc}
\hline
\hline
$m_\sigma$ (MeV)& $\delta_s$ (MeV)& $g_{\omega N}$ & $c_\omega$ & $g_{\rho N}$ & $g_{\sigma\Lambda}$ & $g_{\zeta\Lambda}$\\
\hline
690 & 32.81 & 11.95 & 294.9 & 4.54 & 3.40 & 5.17 \\
\hline
\hline
\end{tabular}
\end{table}

There are two points to be noted in view of the chiral symmetry
in the above Lagrangian.
We omit pseudoscalar meson effects,
and we do not require chiral symmetry in baryon-scalar meson couplings.
In discussing the in-medium modification of the chiral condensates,
the linear representation is more convenient,
where scalar and pseudoscalar mesons appear as chiral partners.
In the mean field treatment with parity fixed single particle baryon states,
the expectation values of pseudoscalar mesons disappear,
and we omit the explicit role of pseudoscalar mesons.
Based on the chiral perturbation (ChPT) theory~\cite{Finelli:2005ni}
and in the extended relativistic chiral mean field model~\cite{Hu:2009zza},
two pion exchange effects were examined
in the relativistic nuclear energy density functional
and they would modify the scalar and vector coupling constants
in a density dependent way.
We here assume that two pion exchange effects are represented
in the coupling constants and the self-interaction terms of scalar and vector
mesons.
In the linear representation,
it is possible to construct an SU(3) chiral symmetric Yukawa coupling term
of scalar and pseudoscalar mesons with baryons~\cite{Papazoglou1998},
but we have to assume baryons transform as nonet
and only D-type coupling appears.
For octet baryons having both D- and F-type Yukawa couplings,
it is necessary to introduce two-types of baryons~\cite{Christos1987},
or to invoke the non-linear
representation~\cite{Papazoglou1999,Schramm2002},
which are out of the scope in this paper.

\section{Finite nuclei and nuclear matter}\label{Results}

\subsection{Nuclear matter and normal nuclei}\label{Subsec:Normal}

\begin{figure}
\PSfig{8.5cm}{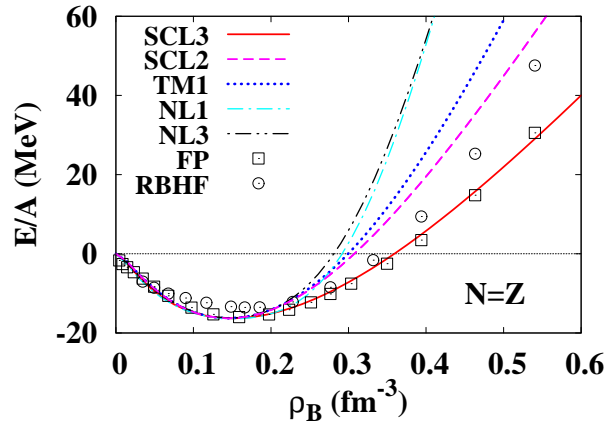}
\caption{(Color online)
	 Energy per nucleon in symmetric nuclear matter
	 as a function of the baryon density.
         Solid, dashed, dotted, dot-dashed, dot-dot-dashed curves
	 show the results in
	 SCL3($m_\sigma = 690$), SCL2~\cite{SCL2}, TM1~\cite{TM1},
	 NL1~\cite{NL1} and NL3~\cite{NL3},
	 respectively.
         Calculated results in variational calculation (FP)~\cite{FP81}
         and RBHF~\cite{RBHF} are also presented
	 by open squares and and open circles, respectively.
        }
\label{fig:SymEOS}       
\end{figure}

First we discuss the EOS of symmetric nuclear matter.
There are three relevant parameters,
$m_\sigma$, $g_{\omega N}$ and $c_\omega$.
For a given $m_\sigma$ value, 
latter two are determined by fitting the saturation point
$(\rho_0, E_0/A) = (0.15~\mbox{fm}^{-3}, -16.3~\mbox{MeV/A})$.
In Fig. \ref{fig:SymEOS},
we show calculated energy per nucleon ($E/A$) in TM1, SCL2,
and the present model (SCL3) with $m_\sigma=690~\mathrm{MeV}$.
When we adopt $m_\sigma\sim700\mbox{MeV}$,
the EOS in SCL3 is considerably softer than those in TM1 and SCL2.
We also find the SCL3 EOS is in good agreement with
the variational calculation results
by Friedman and Pandharipande (FP)~\cite{FP81},
especially at around $\rho_0$.
SCL3 EOS has rather soft incompressibility $K\sim211.0$ MeV and this result
is comparable with the empirical incompressibility $K=210\pm30$ MeV~\cite{Blaizot:1980tw}.

\begin{figure}
\PSfig{8.5cm}{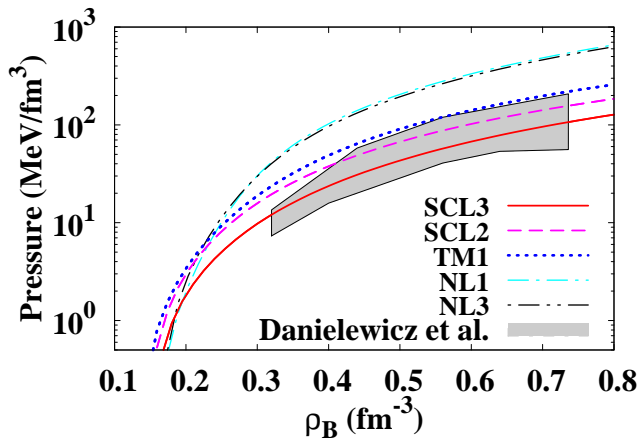}
\caption{(Color online)
	 Same as Fig.~\protect{\ref{fig:SymEOS}}
	 but for the pressure.
	 Shaded area shows the region of pressures consistent
	 with the experimental flow data
	 analyzed by using
	 the Boltzmann equation model~\protect{\cite{Danielewicz:2002pu}}.
}
\label{fig:prs}       
\end{figure}

Nuclear matter EOS at several $\rho_0$ has been probed
in heavy-ion collisions~\cite{Danielewicz:2002pu,Sahu:1999mq,Isse:2005nk}.
In Fig.~\ref{fig:prs},
we show the region of pressures consistent with the experimental flow data
analyzed by using the Boltzmann equation model~\cite{Danielewicz:2002pu}.
Danielewicz, Lacey and Lynch suggested the range of the incompressibility
$167~\mathrm{MeV} \leq K \leq 380~\mathrm{MeV}$ in the density range
$2\rho_0 \leq \rhoB \leq 5\rho_0$~\cite{Danielewicz:2002pu}.
Other theoretical model calculations~\cite{Sahu:1999mq,Isse:2005nk}
also explain flow data at AGS and SPS energies
with $K \simeq 300~\mathrm{MeV}$.
The calculated pressure in SCL3 is consistent with the pressure range
suggested in Ref.~\cite{Danielewicz:2002pu}.

\begin{figure}
\PSfig{8.5cm}{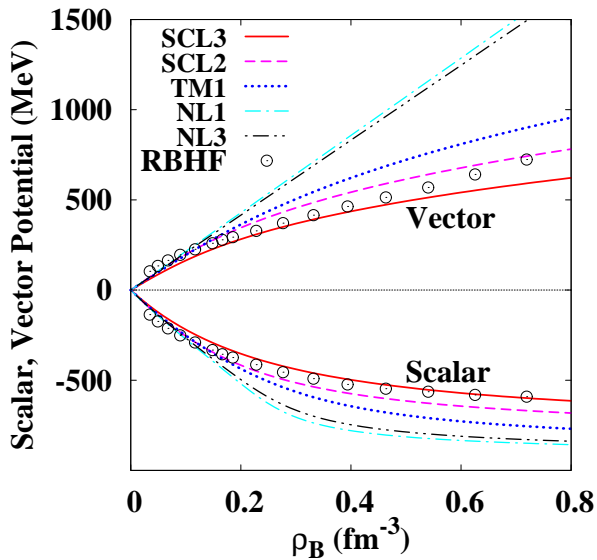}
\caption{(Color online)
	 Same as Fig.~\protect{\ref{fig:SymEOS}}
	 but for the scalar and vector potentials.
}
\label{fig:matpot}       
\end{figure}

The EOS softening is caused by the $\zeta$ meson,
which couples with $\sigma$ through the determinant interaction.
In Fig.~\ref{fig:szmix}, we show the density dependence of
the equilibrium point in $(\varphi_\sigma, \varphi_\zeta)$ plane.
Equilibrium values at $\rhoB/\rho_0 = 0 \sim 5$ are shown with points.
We find that the system evolves along with the valley
where $\zeta$ values are finite.
This $\zeta$ variation reduces the quadratic part of the mesonic
energy density,
while both $\varphi_\sigma$ and $\varphi_\zeta$ contribute
repulsively in the higher order term, $V_{\sigma\zeta}$.
As a result, the energy gain from the scalar mesons is suppressed a little,
and it leads to a smaller vector coupling to reproduce the saturation point.
Cancellation of smaller scalar and vector potentials leads to
a softer EOS in SCL3.

The density dependence of the scalar and vector potentials
in SCL3 are found to be qualitatively consistent
with the RBHF results~\cite{RBHF}
at low densities, $\rhoB < 0.3~\mathrm{fm}^{-3}$.
In Fig.~\ref{fig:matpot}, we show the scalar and vector potentials,
$U^s_N=M^*_N - M_N=-g_{\sigma N}\varphi_\sigma$
and $U^0_N=g_{\omega N}\omega$ in symmetric nuclear matter,
as functions of density in SCL3 in comparison with those
in SCL2, TM1, and RBHF.
Scalar and vector potentials grow almost linearly with $\rhoB$
at very low densities, and they are suppressed at higher densities
in RBHF via the exchange and correlation,
and the relativistic normalization~\cite{RBHF}.
In RMF models, the suppression is caused by the non-linear terms
of $\sigma$ and $\omega$.
The RBHF results lie between SCL2 and SCL3,
and the density dependence at $\rhoB < 0.3~\mathrm{fm}^{-3}$
is well reproduced with SCL3.
Once these potentials are given, similar EOSs are obtained at low densities;
the difference appears from the non-linear terms, whose residual effects
are small at low densities.
At high densities ($\rhoB > 0.3~\mathrm{fm}^{-3}$),
SCL3 gives softer EOS than that in RBHF.
This difference may come from the density dependence of the vector potential.
As shown in Fig.~\ref{fig:matpot},
the vector potential in SCL3 is suppressed more strongly than in RBHF.
Since the $\omega^4$ term is introduced to mimic the density dependence
in RBHF, it may be necessary to include other types of non-linear interaction
terms for vector field
provided that the density dependence in RBHF gives the convergent result
in the hole-line expansion.
 
\begin{figure}
\PSfig{8.5cm}{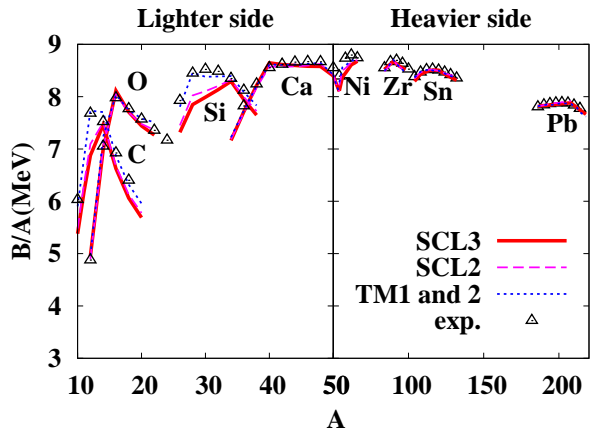}
\caption{(Color online)
         Binding energy per nucleon in normal nuclei from C to Pb isotopes.
         Solid, dashed and dotted curves show calculated results
	 of SCL3, SCL2, and TM
	 (TM2 for C to Ca isotopes and TM1 for Ni to Pb isotopes)
	 models, respectively.
	 Experimental data are shown in open triangles.
         }
\label{fig:bepn}       
\end{figure}

For normal finite nuclei, binding energies per nucleon and charge rms radii are
controlled by two parameters, $m_\sigma$ and $g_{\rho N}$.
We determine them by fitting experimental data of binding energies
and charge rms radii of some stable nuclei
(\Nucl{12}{C}, \Nucl{16}{O},
\Nucl{40}{Ca}, \Nucl{48}{Ca}, \Nucl{58}{Ni} and \Nucl{90}{Zr})
and Sn and Pb isotopes.
In Fig. \ref{fig:bepn},
we show the calculated binding energies per nucleon ($B/A$)
of C, O, Si, Ca, Ni, Zi, Sn and Pb isotopes,
and $B/A$ and charge rms radii for some stable nuclei are shown
in Tables \ref{Table:be} and \ref{Table:crms},
where we also tabulate
NL1~\cite{NL1}, NL3~\cite{NL3},
and non-chiral RMF (TM~\cite{TM1, HT01}) results.
 
From these results, we find that SCL3 RMF model well describes
the bulk properties of normal nuclei with the parameters shown
in Table ~\ref{tab:params}, 
especially those of Sn and Pb isotopes which we can treat as spherical.
Since $B/A$ in these nuclei apparently reflects the character 
of EOS around $\rho_0$, our choice of the saturation point
and adopted values of parameters
($m_\sigma = 690\mbox{MeV}$, $g_{\rho N} = 4.54$) is
appropriate in explaining these data.

\begin{table}
\begin{minipage}{0.48\textwidth}
\centering
\caption{Experimental and theoretical binding energies and of stable nuclei.
         The results obtained from the SCL3 model are compared
	 with those obtained from
	 TM1,~\protect{\cite{TM1}}
	 TM2,~\protect{\cite{TM1}}
	 NL1,~\protect{\cite{NL1}}
	 NL3,~\protect{\cite{NL3}}
	 and SCL2~\protect{\cite{SCL2}} models
         and with experimental data.
}
\label{Table:be}
\vspace{10pt}
\begin{tabular}{cccccccc}
\hline
\hline
        \multicolumn{8}{c}{$B/A$ (MeV)} \\
\hline
Nucleus      &exp.&SCL3&SCL2 &TM1 &TM2 &NL1 &NL3 \\
\hline
${}^{12}$C   &7.68&6.91&7.09& -  &7.68& -  & -  \\
${}^{16}$O   &7.98&8.11&8.06& -  &7.92&7.95&8.05\\
${}^{28}$Si  &8.45&7.85&8.02& -  &8.47&8.25& -  \\
${}^{40}$Ca  &8.55&8.64&8.57&8.62&8.48&8.56&8.55\\
${}^{48}$Ca  &8.67&8.58&8.62&8.65&8.70&8.60&8.65\\
${}^{58}$Ni  &8.73&8.44&8.54&8.64& -  &8.70&8.68\\
${}^{90}$Zr  &8.71&8.67&8.69&8.71& -  &8.71&8.70\\
${}^{116}$Sn &8.52&8.51&8.51&8.53& -  &8.52&8.51\\
${}^{196}$Pb &7.87&7.84&7.87&7.87& -  &7.89& -  \\
${}^{208}$Pb &7.87&7.87&7.87&7.87& -  &7.89&7.88\\
\hline
\hline
\end{tabular}
\end{minipage}
\end{table}
\begin{table}
\begin{minipage}{0.48\textwidth}
\centering
\caption{Same as Table~\ref{Table:be} but for charge rms radii.}
\label{Table:crms}
\vspace{10pt}
\begin{tabular}{cccccccc}
        \hline
        \hline
        \multicolumn{8}{c}{charge rms radius (fm)} \\
        \hline
Nucleus     &exp.&SCL3&SCL2 &TM1 &TM2 &NL1 &NL3 \\
        \hline
${}^{12}$C  &2.46&2.47&2.43& -  &2.39& -  & -  \\
${}^{16}$O  &2.74&2.63&2.62& -  &2.67&2.74&2.73\\
${}^{28}$Si &3.09&3.06&3.04& -  &3.07&3.03& -  \\
${}^{40}$Ca &3.45&3.43&3.44&3.44&3.50&3.48&3.47\\
${}^{48}$Ca &3.45&3.46&3.46&3.45&3.50&3.44&3.47\\
${}^{58}$Ni &3.77&3.78&3.77&3.76& -  &3.73&3.74\\
${}^{90}$Zr &4.26&4.26&4.27&4.27& -  &4.27&4.29\\
${}^{116}$Sn&4.63&4.61&4.62&4.61& -  &4.61&4.61\\
${}^{196}$Pb& -  &5.48&5.48&5.47& -  &5.47& -  \\
${}^{208}$Pb&5.50&5.54&5.54&5.53& -  &5.57&5.58\\
        \hline
        \hline
\end{tabular}
\end{minipage}
\end{table}

\subsection{$\Lambda$ hypernuclei}\label{Subsec:Lambda}

From the very early stage~\cite{Glendenning,BW77, BB81},
RMF models have been applied to 
$\Lambda$ hypernuclei.
Later, in Ref.~\cite{TM1Lam,MJ94}, 
tensor type couplings between vector meson and $\Lambda$ were introduced
so as to explain the small \textit{ls} splitting.
This problem was also examined from the view of baryon-meson
density dependent coupling~\cite{DDRHL}.
In addition to $\sigma, \omega$ and $\rho$ mesons,
hidden strange meson fields ($\zeta$ and $\phi$) were introduced
in order to explain additional hyperon-hyperon interaction~\cite{Schaffner}.
In these works, hyperon-meson coupling constants are determined based on
the flavor SU(3) or flavor-spin SU(6) symmetry,
$\Lambda$ hypernuclear single particle energies,
hyperon potential depths in nuclear matter,
and suggestions from $YN$ g-matrix.
In Refs.~\cite{BW77,BB81},
$x \equiv g_{\sigma\Lambda}/g_{\sigma N}\simeq g_{\omega\Lambda}/g_{\omega N}\simeq 1/3$
is adopted from discussions of $\pi$, $\rho$ and $\omega$ exchanges~\cite{BW77}
or by fitting single particle levels of $\Lambda$ hypernuclei~\cite{BB81},
while in Ref.~\cite{Glendenning},
the ratio is set to be $x=2/3$ from light quark counting arguments.
In Refs.~\cite{MJ94,Schaffner},
the vector meson-hyperon couplings are fixed
from the SU(6) relation or the additive quark model results,
{\em e.g.} 
$g_{\omega\Lambda}=g_{\omega\Sigma}=2g_{\omega\Lambda}=2/3 g_{\omega N}$,
while the scalar meson-hyperon couplings are determined
from the $\Lambda$ hypernuclear single particle energies
or the hyperon potential depths in baryonic matter.
In Ref.~\cite{TM1Lam},
several sets of parameters are compared
in the range $g_{\omega\Lambda}/g_{\omega N}=0.18-0.63$,
and it is concluded that the $\Lambda$ single particle energies are not
enough to determine the meson-$\Lambda$ coupling constants.

Since the experimental information on $YN$ interaction is limited,
it is valuable to study the hypernuclear systems with the RMF models including
the chiral potential which can explain normal nuclear property.
In previous studies~\cite{Papazoglou1998,Papazoglou1999},
chiral SU(3) symmetric RMF models are proposed.
In the linear~\cite{Papazoglou1998} and non-linear~\cite{Papazoglou1999}
representation, various types of meson Lagrangian are compared.
These models describe the normal nuclear properties
very well~\cite{Papazoglou1999,Schramm2002},
while the hypernuclear properties in these models are not
satisfactory from a phenomenological point of view.
Especially $\Sigma$ and $\Xi$ hyperons are considered to feel
repulsive~\cite{SigmaPot} and weakly attractive~\cite{XiPot} potentials,
but these features are not explained yet.
This may be suggesting the importance
of other types of meson-hyperon couplings other than the Yukawa coupling
or the SU$_f$(3) breaking effects.

In the present work, we study single- and double-$\Lambda$ hypernuclei
in the SCL3 RMF model, which has already shown to work well in normal nuclei
and nuclear matter as demonstrated in the previous subsection.
We adopt the $\mathrm{SU}_f(3)$ relation for vector coupling,
and scalar meson-$\Lambda$ Yukawa coupling constants are chosen to
fit the $\Lambda$ separation energies of single $\Lambda$ hypernuclei
and the $\Lambda\Lambda$ bond energy in $^6_{\Lambda\Lambda}$He.
In this treatment, the scalar meson-$\Lambda$ coupling terms
do not necessarily preserve the chiral SU(3) symmetry,
but this may be an appropriate prescription at present
because of the phenomenological problems
in the chiral SU(3) RMF mentioned above.
We omit the vector meson-hyperon tensor couplings,
which does not affect the EOS in the mean field treatment.

In single- and double-$\Lambda$ hypernuclei,
we have four adjustable parameters,
$g_{\sigma\Lambda}$, $g_{\zeta\Lambda}$, $g_{\omega\Lambda}$ and
$g_{\phi\Lambda}$.
Here, we assume
that the vector couplings obey
the lowest order SU$_f$(3) symmetric relation~\cite{Dover:1985ba,M93},
\begin{align}
\mathcal{L}_{\mathrm{BV}}
= \sqrt{2}\{& g_s\,\tr\sbrace{M_v} \tr\sbrace{\bar{B}B}
              + g_D\,\tr\sbrace{\bar{B}\mbrace{M_v, B}} \nonumber\\
		    & + g_F\,\tr\sbrace{\bar{B}\bbrace{M_v, B}}\}\nonumber\\ 
= \sqrt{2}\{& g_s\,\tr\sbrace{M_v} \tr\sbrace{\bar{B}B}
              + g_1\,\tr\sbrace{\bar{B}M_vB} \nonumber\\
		    & + g_2\,\tr\sbrace{\bar{B}BM_v}\}\ ,
\label{SU3coupling}
\end{align}
where $g_D = (g_1 + g_2)/2$, $g_F = (g_1 - g_2)/2$,
and the Lorentz indices and gamma matrices are assumed.
In this form of vector coupling,
$g_{\omega\Lambda}$ and $g_{\phi\Lambda}$ are already fixed by
the vector coupling constants with nucleons, $g_{\omega N}$ and $g_{\rho N}$,
as
\begin{equation}
g_{\omega\Lambda} = \frac{5}{6}g_{\omega N} - \frac{1}{2}g_{\rho N},\;\;
g_{\phi\Lambda} = \frac{\sqrt{2}}{6}\sbrace{g_{\omega N} + 3g_{\rho N}}\ .
\label{LambdaVM}
\end{equation}
Scalar coupling constants, 
$g_{\sigma\Lambda}$ and $g_{\zeta\Lambda}$,
are then tuned to reproduce existing data of $S_\Lambda$
and the $\Lambda\Lambda$ bond energy, $\Delta B_{\Lambda\Lambda}$,
observed in the Nagara event, \DLNucl{6}{He}~\cite{Nagara}.

\begin{figure}
\centering
\PSfig{8.5cm}{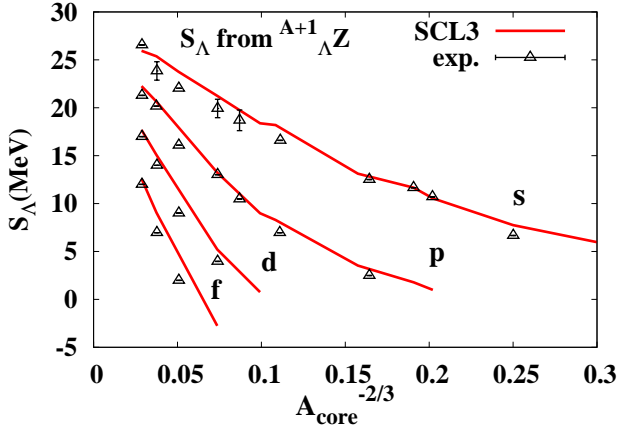}
\caption{(Color online)
         Separation energy of $\Lambda$.Parameters are determined by fitting
         $S_\Lambda$ of \LNucl{13}{C} and \LNucl{12}{C}.
         $S_\Lambda$ of $p$, $d$ and $f$ are wait-averaged considering the state
         numbers in each levels.}
\label{fig:sLambda}       
\end{figure}

In Fig. \ref{fig:sLambda},
we show the calculated results of $S_\Lambda$
with the parameter set in Table ~\ref{tab:params}.
Here, we evaluate the zero-point kinetic energy, $E_\mathrm{ZPE}$,
with a harmonic-oscillator wave function 
as $E_\mathrm{ZPE} = \frac{3}{4}\cdot 41A_\mathrm{core}^{-1/3}~\mathrm{MeV}$.
The results of the $S_\Lambda$ for $p$, $d$ and $f$ levels
are the weight-averaged ones of the spin-orbit partners.

\begin{figure}
\centering
\PSfig{8.5cm}{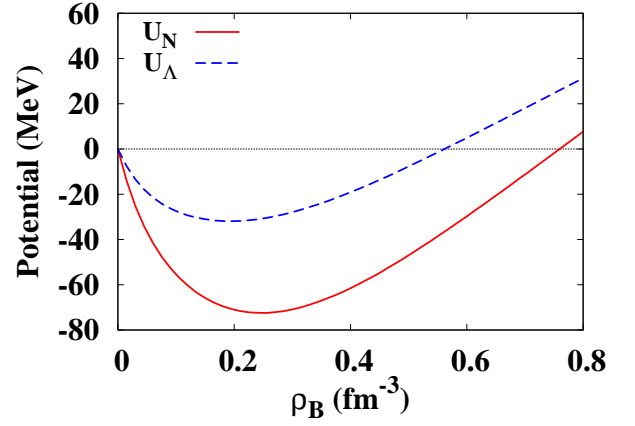}
\caption{(Color online)
	 Single particle potential of nucleon and $\Lambda$.
	 Solid and dashed curves show the potential of nucleon and $\Lambda$,
	 respectively.
         }
\label{fig:ULam}       
\end{figure}

The scalar potential of $\Lambda$ is given
in the form of linear combination of the coupling constants
and chiral condensates,
\begin{align}
U^s_\Lambda(\rhoB)
=-[g_{\sigma\Lambda}\varphi_\sigma(\rhoB)
+ g_{\zeta\Lambda}\varphi_\zeta(\rhoB)]\ .
\end{align}
In Fig.~\ref{fig:ULam}, we show the single particle potential for $\Lambda$,
defined as the sum of the scalar and vector potentials,
$U_\Lambda=U^s_\Lambda+U^{0}_\Lambda$,
where the temporal component of the vector potential $U^0_\Lambda$ is defined
in Eq.~(\ref{Eq:VectorPot}).
This sum roughly corresponds to the Schr\"odinger equivalent potential
for $\Lambda$, $U^{(\mbox{\scriptsize SEP})}_\Lambda=U^s_\Lambda+(E/M_\Lambda)
U^0_\Lambda$.
As in the previous studies~\cite{Dover:1985ba,Millener:1988hp,Bando:1990yi},
$U_\Lambda$ amounts to be around $-30~\mathrm{MeV}$ at $\rho_0$.
Calculated $S_\Lambda$ values are very similar
as far as the scalar potentials at $\rho_0$ are the same.

The $ls$ splitting between $p_{1/2}^\Lambda$ and $p_{3/2}^\Lambda$
in \LNucl{13}{C} is calculated to be 900 keV in the present treatment.
This result is larger than the observed small $ls$ splitting~\cite{A01}. 
A small $ls$ splitting would be obtained by including the tensor type couplings
between meson and $\Lambda$ hyperon~\cite{TM1Lam},
since this coupling directly corresponds 
to $ls$ force when we translate the Dirac equation in RMF
into a Schr\"odinger equivalent form. 
This coupling, however, reduces only the $ls$ splitting and does not change
the average location of $ls$ partners.
Thus, we suppose that determined $g_{\sigma\Lambda}$ and $g_{\zeta\Lambda}$
are not affected dramatically.
Two pion exchange anti-$ls$ force is also suggested
as the origin of the small $ls$ splitting and examined
by introducing density dependent coupling between meson and baryon~\cite{DDRHL}
and in the in-medium chiral SU(3) dynamics~\cite{Finelli:2007wm}.

\begin{figure}
\PSfig{8.5cm}{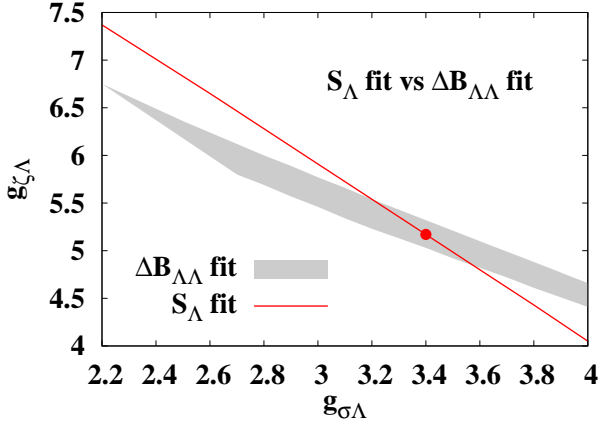}
\caption{(Color online)
$g_{\sigma\Lambda}$ and $g_{\zeta\Lambda}$ suggested from experimental data.
Solid line shows the relation between
$g_{\sigma\Lambda}$ and $g_{\zeta\Lambda}$
determined by fitting $S_\Lambda$ of \LNucl{12}{C} and \LNucl{13}{C},
and gray shaded area shows the region
which explains the experimental $\Delta B_{\Lambda\Lambda}$
value of \DLNucl{6}{He} within the error.
The filled point represents
the pair of $g_{\sigma\Lambda}$ and $g_{\zeta\Lambda}$,
which explains $S_\Lambda$ and the central value of the experimental bond energy
($\Delta B_{\Lambda\Lambda}$) simultaneously,
and is adopted in the present work.
}\label{fig:delBll-map}       
\end{figure}

In Fig.~\ref{fig:delBll-map},
we show the relation of $g_{\sigma\Lambda}$ and $g_{\zeta\Lambda}$
determined by fitting $S_\Lambda$ of \LNucl{12}{C} and \LNucl{13}{C}.
This relation is roughly evaluated as
$\widetilde{g}_{\sigma\Lambda}=g_{\sigma\Lambda}+g_{\zeta\Lambda}/2 \simeq 6$.
All the parameter sets on the solid line can reproduce $S_\Lambda$
of various single $\Lambda$ hypernuclei similarly to the one shown
in Fig.~\ref{fig:sLambda}.
We also examine $\Delta B_{\Lambda\Lambda}$ in this parameter plane
and find that the parameter sets in the gray shaded area
explain the experimental $\Delta B_{\Lambda\Lambda}$ value of \DLNucl{6}{He}
within the error.
Combining these results, we obtain the set of coupling constants as
$(g_{\sigma\Lambda}$, $g_{\zeta\Lambda})=(3.40, 5.17)$
shown in Table ~\ref{tab:params},
which explains $S_\Lambda$ and the central value of the experimental bond energy
($\Delta B_{\Lambda\Lambda}$) simultaneously.

The above set of coupling constants
$(g_{\sigma\Lambda}$, $g_{\zeta\Lambda})$
deviates from a naive estimate,
$g_{\sigma\Lambda}/g_{\sigma N}=2/3$
and 
$g_{\sigma\Lambda}/g_{\zeta\Lambda}=\sqrt{2}$.
While the $\sigma\Lambda$ coupling is small,
the scalar potential for $\Lambda$ is around $2/3$ of that for nucleons,
and it is still dominated by $\sigma$.
The additional scalar potential comes from $\zeta$,
which does not couple with nucleons directly
but appears from the $\sigma\zeta$ mixing
generated by the KMT term in the Lagrangian.
Since $\varphi_\zeta$ evolves as $\varphi_\zeta \simeq \varphi_\sigma/2$
in symmetric nuclear matter as seen in Fig.~\ref{fig:szmix},
the scalar potential of $\Lambda$ behaves as
$U^s_\Lambda \simeq - \widetilde{g}_{\sigma\Lambda}\,\varphi_\sigma$.
The effective coupling $\widetilde{g}_{\sigma\Lambda}$
is around 2/3 of $g_{\sigma N}$,
then the smaller $\Lambda$ scalar potential from $\sigma$
is compensated by the $\zeta$ meson.
The obtained $g_{\sigma\Lambda}$ is close to 1/3 of $g_{\sigma N}$,
as suggested from the two pion exchange~\cite{BW77}.
These observations may be suggesting the importance of pions and KMT term
at finite densities.

\subsection{Neutron star matter}\label{Subsec:NS}

In previous subsections,
we have fixed all parameters for nucleon and $\Lambda$
by fitting the empirical saturation point of symmetric nuclear matter,
and the binding energies of normal nuclei and $\Lambda$ hypernuclei.
Now we apply the SCL3 RMF model to neutron star (NS) matter.

In NS matter, we require neutrinoless $\beta$--equilibrium
and charge neutrality condition,
\begin{gather}
\mu_i = b_i\mu_B - q_i\mu_e\ ,\\
\rho_c =  \sum_B q_B\rho_v^{(B)} + \sum_l q_l\rho_v^{(l)} = 0\ .
\label{betaEqu}
\end{gather}
The first equation relate the chemical potentials as,
$\mu_n = \mu_p+\mu_e = \mu_\Lambda$ and $\mu_e = \mu_\mu$.
Under this equilibrium condition,
we calculate the energy density ($\varepsilon$) and pressure ($P$),
and we solve the Tolman--Oppenheimer--Volkoff(TOV) equation,
\begin{equation}
\diff{P}{r} = - \frac{[P(r) + \varepsilon(r)][M(r)+4\pi r^3P(r)]}{r(r-2M(r))}
\ .
\label{TOV}
\end{equation}
where $M(r)$ denotes the mass inside the radius $r$.
Here, we neglect nuclear crust for simplicity.
 
\begin{figure}
\PSfig{8.5cm}{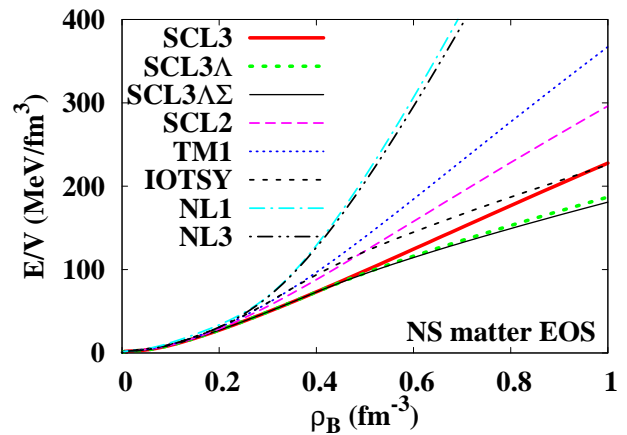}
\caption{(Color online)
         Energy per baryon in neutron star matter.
	 Solid, short-dashed, bold-dashed, dotted, thin-dashed,
	 dot-dashed, dot-dot-dashed curves show the results of
         SCL3, SCL3$\Lambda$, SCL2, TM1, IOTSY, NL1 and NL3 results,
	 respectively.
}\label{fig:NSEOS}       
\end{figure}

In Fig. \ref{fig:NSEOS}, we show the NS matter EOS.
We compare the results of SCL3 RMF model only with nucleons (SCL3)
and with $\Lambda$ hyperons (SCL3$\Lambda$).
We also show the NS matter EOS in SCL2~\cite{SCL2},
TM1~\cite{TM1}, NL1~\cite{NL1}, NL3~\cite{NL3} and IOTSY~\cite{IOTSY},
where $\Lambda$ hyperons are not taken into account except for IOTSY.
The IOTSY RMF model is based on TM1 model and
includes all octet hyperons ($\Lambda, \Sigma, \Xi$).
We find that the NS matter EOS in SCL3 is softer
than those in other RMF EOSs without hyperons.
Especially, when we include $\Lambda$,
$\Lambda$ hyperons appear at around 2$\rho_0$
and NS matter EOS including $\Lambda$ hyperons becomes further softer.

\begin{figure}
\PSfig{8.5cm}{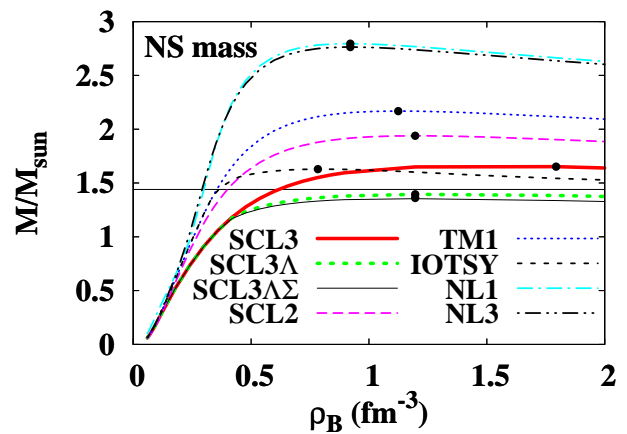}
\caption{(Color online)
	 Same as Fig.~\ref{fig:NSEOS} but for
	 the neutron star mass as a function of the central density.
}\label{fig:NSmass}       
\end{figure}

In Fig. \ref{fig:NSmass}, 
we show the results of neutron star mass as a function of the central density
with SCL3, SCL3$\Lambda$, SCL2, TM1, IOTSY, NL1 and NL3 models.
While the EOS in SCL3 is much softer than other EOSs,
calculated maximum NS mass in SCL3 (without hyperon) is $1.65 M_\odot$,
which exceeds the precisely observed NS mass, 1.44 $M_\odot$~\cite{NSmass}.
When $\Lambda$ hyperons are included,
the maximum mass is calculated to be $1.40 M_\odot$ with SCL3$\Lambda$
and underestimates the observed mass.

This underestimate is caused by the softer EOS,
particularly in the high $\rhoB$ region.
It is suggested that extra repulsion coming from three-baryon interactions
or string-junction model~\cite{Tamagaki} 
which are repulsive for all baryon universally
are needed to surpass the known NS mass data
in non-relativistic calculations~\cite{YNT91,Baldo,Nishizaki}.
In RMF models, the EOS of nuclear matter is stiff enough,
and extra repulsion is not generally required to support $1.44~M_\odot$.
In SCL3 and SCL3$\Lambda$, however,
the incompressibility is as small as the empirical value
and the pressure at high density region is compatible
with the estimate in heavy-ion collisions~\cite{Danielewicz:2002pu}.
Thus, we encounter the same problem as in the non-relativistic calculations
and have to consider additional repulsions
which have a large effect at high densities.
One of the candidates of this extra repulsion may come from
the $\sigma\omega$ coupling,
such as in the term of $\sigma^2\omega^2$~\cite{Boguta,Ogawa2004}.
The $\sigma\omega$ coupling results in reducing of the vector
meson mass at high densities, and is found to give very stiff EOS
when combined with the linear $\sigma$ model.
Inclusion of such coupling may stiffen EOS in high $\rhoB$ region
after re-fitting experimental data
and it may solve the underestimation of the maximum mass of NS.

\begin{figure*}[t]
\centering
\PSfig{17.5cm}{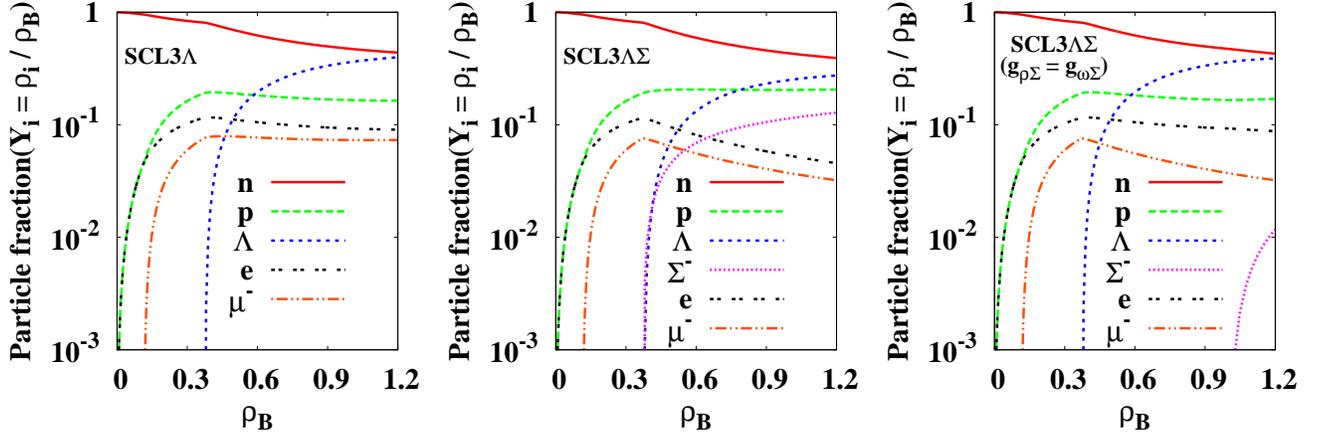}
\caption{(Color online)
         Calculated particle fraction $Y_i$ of $n$, $p$, $\Lambda$, $\Sigma^-$,
	 $e^-$, and $\mu^-$.
	 The other $\Sigma$ hyperons ($\Sigma^0$ and $\Sigma^+$) do not emerge
	 in this baryon density range.
	 }
\label{fig:NSY_B}
\end{figure*}

Now we examine the effects of $\Sigma$ hyperons in neutron star matter.
In neutron star matter, it was believed that the $\Sigma^-$ baryon
would appear at the lowest density among the hyperons
provided that the potential for $\Sigma$ baryons in symmetric nuclear matter
is similar to that for $\Lambda$~\cite{Glendenning,Schaffner}.
The strength of $\Sigma$-nucleus optical potential have been studied
from the atomic shifts of $\Sigma^-$~\cite{B94,Mares1995},
which is sensitive to the attraction in $\Sigma^-$-nucleus potential
at nuclear surface.
In the inner region of nuclei,
the analysis of the $\Sigma^-$ quasi free production spectra
from $(\pi^-, K^+)$ and $(K^-,\pi^+)$ reactions
have yielded that repulsive $\Sigma$-nucleus potential
is favored~\cite{SigmaPot}.
From these point of view,
we employ the repulsive $\Sigma$ potential~\cite{TMMO07}
which is suggested from SU$_f$(3) relation, Eq.~(\ref{SU3coupling}).
Coupling constants, $g_{\omega\Sigma}$ and $g_{\phi\Sigma}$ are given as
\begin{equation}
	g_{\omega\Sigma} = \frac12\sbrace{g_{\omega N} + g_{\rho N}},\ 
	g_{\phi\Sigma}   = \frac{\sqrt{2}}{2}\sbrace{g_{\omega N} - g_{\rho N}}.
	\label{SUf3S}
\end{equation}
From the SU$_f$(3) relation, $g_{\rho\Sigma}$ should be equal to
$g_{\omega\Sigma}$,
\begin{equation}
g_{\rho\Sigma}^\mathrm{SU(3)}=g_{\omega\Sigma}\simeq 2 g_{\rho N}
\ .
\end{equation}
In order to explain the atomic shift data of $\Sigma^-$, however,
we need to adopt a smaller value of $g_{\rho\Sigma}$.
In an RMF model fit, Mares {\em et al.} showed that 
both of Si and Pb atomic shift data~\cite{SigmaAtom}
are well explained by adopting $g_{\rho\Sigma}/g_{\rho N} \simeq 2/3$
~\cite{Mares1995}.
With the present Lagrangian,
we have fitted the $\Sigma^-$ atomic shift data
and have obtained the values
$g_{\rho\Sigma}=1.97$ ($g_{\rho\Sigma}/g_{\rho N}=0.434$)
and $g_{\sigma\Sigma}=3.16$
under the assumption of the na\"ive quark counting coupling ratio
for $\sigma$ and $\zeta$, $g_{\sigma\Sigma}=\sqrt{2}g_{\zeta\Sigma}$.
The calculated EOS, mass of neutron star, and particle fraction
$Y_i = \rho_{i}/\rho_{B}$ with $\Sigma$ hyperons
are shown in Figs.~\ref{fig:NSEOS}, ~\ref{fig:NSmass}, and
in the middle panel of Fig.~\ref{fig:NSY_B},
respectively.
In neutron star matter,
$\Sigma^-$ starts to emerge around $\rho_B \sim 0.4$ as a substitute of leptons
because of its negative charge.
Other $\Sigma$ hyperons ($\Sigma^0, \Sigma^+$)
do not appear even at $10\rho_0$ because of
the repulsive potential and the negative charge chemical potential.
The calculated EOS and maximum mass of neutron star are not affected much.
The repulsive potential suppresses the effects of $\Sigma$ baryons
compared to those of $\Lambda$, which plays a decisive role
as a substitute of the dominant component, $n$.

It is interesting to find that the starting density of $\Sigma^-$ hyperon
($\rhoB \simeq 0.4~\mathrm{fm}^{-3}$) is much lower than those in previous
studies~\cite{Balberg:1997yw,Sahu:2001um,SchaffnerBielich:2008kb,IOTSY},
which also adopt repulsive $\Sigma$ potential in symmetric nuclear matter.
In these works, $\Sigma^-$ appears at much higher density.
The main difference in the present work is the coupling strength
with the $\rho$ meson. 
In Refs.~\cite{SchaffnerBielich:2008kb,IOTSY},
the SU(3) value ($g_{\rho\Sigma}/g_{\rho N}=2$) is adopted
and the repulsive interaction from $\rho$ is strong
in high density neutron star matter,
while the present coupling ($g_{\rho\Sigma}/g_{\rho N}=0.434$)
is much smaller and the repulsive potential from $\rho$ is weaker.
In order to demonstrate this point, we show the particle fraction results
with $g_{\rho\Sigma}/g_{\rho N}=2$ in the right panel of Fig.~\ref{fig:NSY_B}.
We find that $\Sigma^-$ appears only at $\rhoB \gtrsim 1~\mathrm{fm}^{-3}$,
which is qualitatively consistent with previous
works~\cite{Balberg:1997yw,Sahu:2001um,SchaffnerBielich:2008kb,IOTSY},

One of the problems in the present SCL3 is that
the $c_\omega$ value ($c_\omega=294.9$) is larger than those in
TM1 ($c_\omega=71.3075$) and SCL2 ($c_\omega=200$) models.
The potential term of $c_\omega\omega^4/4$ strongly
suppresses $\omega$ meson field especially at high $\rhoB$~\cite{TM1}.
The lower $c_\omega$ value an RMF model has,
the higher neutron star maximum mass the EOS shows
as seen in Figs.~\ref{fig:NSEOS} and \ref{fig:NSmass}.
When we reduce this parameter by changing $m_\sigma$ value and re-fixing
all the parameter in the way as we discussed,
calculated maximum mass of neutron star should be 1.8 $M_\odot$
on the condition of $m_\sigma = 725$MeV and $c_\omega = 75.66$.
With this choice, however,
we cannot reproduce the binding energies of Sn and Pb isotopes.
Thus, in addition to chiral potential,
the form and strength of vector meson potential is also important
and should be investigated further.

\section{Summary and discussion}
\label{Summary}

In this paper, we have proposed a chiral SU(3) symmetric RMF (SCL3 RMF) model,
and examined its properties
in nuclear matter, normal nuclei, $\Lambda$ hypernuclei
and neutron star matter.
We adopt a logarithmic chiral SU(3) potential,
as the energy density as a function of $\sigma$ at $\rhoB=0$,
derived in the strong coupling limit of lattice QCD~\cite{SCL}.
The Kobayashi-Maskawa-'t Hooft (KMT) determinant interaction
term~\cite{KM1970,tHooft1976} is also introduced
in order to take account of the $\mathrm{U}_A(1)$ anomaly.
Since the chiral symmetry relates the condensates and hadron masses,
the number of parameters are reduced by introducing this symmetry.
After fitting $\pi, K, \mathrm{f}_0(980)$ masses
together with $f_\pi$ and $f_\zeta$,
we have only one free parameter, $m_\sigma$, in the vacuum part.
Under the assumptions that the nucleon mass are fully generated
by the chiral condensate ($M_N = g_{\sigma N}f_\pi$)
and that the nucleon does not couple with $\bar{s}s$ mesons,
we determine four parameters relevant to normal nuclei
($m_\sigma, g_{\omega N}, g_{\rho N}, c_\omega$),
by fitting the empirical and experimental data of
symmetric nuclear matter saturation point,
binding energies and size of normal nuclei.
For $\Lambda$ hypernuclei, we assume that the vector couplings obey
the lowest order $\mathrm{SU}_f(3)$ symmetric relation~\cite{Dover:1985ba,M93},
and the remaining two parameters ($g_{\sigma\Lambda}, g_{\zeta\Lambda}$)
are determined by fitting the experimental data
of the separation energies of single-$\Lambda$ hypernuclei
and the $\Lambda\Lambda$ bond energy in the double-$\Lambda$ hypernucleus,
${}^{~\,6}_{\Lambda\Lambda}\mathrm{He}$.

We find that the SCL3 model
well describes the symmetric nuclear matter properties
and the bulk properties of normal nuclei:
The equation of state (EOS) is found to be softened
by the $\sigma\zeta$ coupling generated by the KMT interaction,
and the incompressibility of symmetric nuclear matter is found to be
$K \simeq 210~\mathrm{MeV}$,
which is consistent with the empirical value,
$K=210 \pm 30~\mathrm{MeV}$~\cite{Blaizot:1980tw}.
The EOS around $\rho_0$ is in agreement with
the results of variational calculations~\cite{FP81},
and the pressure in the density region of $2\rho_0\leq\rhoB\leq 5\rho_0$
is in agreement with the estimates
from heavy-ion collision data~\cite{Danielewicz:2002pu}.
The density dependence of the vector potential is close to that
in the relativistic Br\"uckner-Hartree-Fock (RBHF) calculation~\cite{RBHF}
at low densities, $\rhoB<3\rho_0$.
The binding energies of normal nuclei from C to Pb isotopes are 
reasonably well explained except for the $jj-$closed shell nuclei.
Single- and double-$\Lambda$ hypernuclei are also well described.
Separation energies of $\Lambda$ in single-$\Lambda$ hypernuclei,
$S_\Lambda$, are mainly determined by the potential depth,
and the $\Lambda\Lambda$ bond energy depends
on the $\zeta\Lambda$ coupling $g_{\zeta\Lambda}$ more strongly
than on the $\sigma\Lambda$ coupling $g_{\sigma\Lambda}$.

The calculated maximum neutron star mass in the present SCL3 RMF model
underestimates the observed neutron star mass $1.44M_\odot$.
This underestimate would originate from the soft EOS of nuclear matter
at high densities.
The vector potential is suppressed more strongly from the linear behavior,
$\omega \sim g_{\omega N}\rhoB/m_\omega^2$, than the RBHF results,
and the EOS of symmetric matter is softer than the results in
the variational calculation~\cite{FP81} and RBHF~\cite{FP81} at high densities.
The suppression of the vector potential is caused by the $\omega$
self-interaction term, $-c_\omega\omega^4/4$, whose coefficient is large
in SCL3 compared with previous RMF models with this term.
Since this term is introduced phenomenologically to simulate the suppression
of the vector potential in RBHF~\cite{TM1},
it would be necessary to introduce other types of coupling,
such as the scalar meson-vector meson coupling, 
$\sigma^2\omega^2$~\cite{Boguta,Ogawa2004}.
The scalar-vector coupling acts to modify
the in-medium vector meson mass~\cite{NJL_PRep,N06}.
It may be interesting to invoke the results of strong coupling lattice QCD 
with finite coupling effects, where the plaquette contribution is found to
generate the vector potential~\cite{NLO}.
Another possibility is to introduce the repulsive three-baryon force,
which is widely adopted
in non-relativistic theories~\cite{Baldo,FP81,APR,Nishizaki}.

In this paper, we examine how $\Lambda$ and $\Sigma$ hyperons affect
the neutron star matter EOS based on experimental data.
The isospin dependence of $\Sigma$ potential in nuclear matter
is found to be important for the composition at high densities.
$\Xi$ hyperons are not included because the data
are not enough to constrain the potential~\cite{XiPot}.
Since $\Xi$ hyperons may further soften EOS at high densities,
it is necessary to find the mechanism of re-stiffening at high densities
in order to construct reliable and chiral SU(3) symmetric EOS
including all these hyperons.
Explicit role of pions~\cite{Finelli:2005ni,Hu:2009zza,Pion}
is another important subject
to study in terms of relativistic nuclear many-body problems.
The $ls$-like potential~\cite{Finelli:2007wm,PionLS} from pion exchange
would improve the binding energies of $jj$-closed shell nuclei.
In addition, tensor suppression
may be play a critical role in EOS at higher densities.

\begin{acknowledgments}
We would like to thank Professor Avraham Gal
and Jiri Mares for useful discussions.
This work was supported in part by KAKENHI from MEXT and JSPS
under the grant numbers,
	17070002,		
	19540252		
and
	20$\cdot$4326,  
Global COE Program
"The Next Generation of Physics, Spun from Universality and Emergence",
and the Yukawa International Program for Quark-hadron Sciences (YIPQS).
Discussions during the YIPQS international workshop on
"New Frontiers in QCD 2010",
were useful to complete this work.
\end{acknowledgments}

\appendix
\section{The masses of scalar and pseudoscalar mesons\label{AppA}}
We show the formulae of the scalar and pseudoscalar meson masses
except for $\sigma$, $\zeta$, $\pi$ and $K$,
which we have already shown in Sec.~\ref{SCL3}.
When we adopt expectation value of each mesons and mass values of $\pi$, $K$
and $f_0$ as constraints,
mass of $a_0$, $\eta$ and $\eta'$ can be represented
as a function of parameter $m_\sigma$.
These masses can be read as
\begin{gather}
m_{a_0}^2 = b' + \frac{2a'}{f_\pi^2} + 2d'\fz\,\\
m_\kappa^2
	= b' + \frac{\sqrt{2}a'}{\fpi \fz'} + \sqrt{2}d'f_\pi\ ,\\
m_{\eta}^2 = b' - \frac{2a'}{\fpi^2}+2d'\fz\ ,\\
m_{\eta_s}^2 = b' - \frac{a'}{{\fz'}^2}\ ,\\
\xi_{\eta\eta_s} = 2d'\fpi\ ,\\
M_\eta^2 = \frac{\sbrace{m_{\eta}^2 + m_{\eta_s}^2}
  - \sqrt{\sbrace{m_{\eta}^2 - m_{\eta_s}^2}^2 + 4\xi_{\eta\eta_s}^2}}{2}\ ,\\
M_{\eta'}^2  = \frac{\sbrace{m_\eta^2 + m_{\eta_s}^2}
  + \sqrt{\sbrace{m_\eta^2 - m_{\eta_s}^2}^2 + 4\xi_{\eta\eta_s}^2}}{2}\ .
\end{gather}
where we use same parameters, such as $a'$, $b'$ and $d'$, defined
in Sec.~\ref{SCL3}.
We have the mixing term of $\eta$ and $\eta_s$ mesons,
thus one have to diagonalize their mass matrix
to obtain vacuum masses.
We tabulate calculated masses as functions of $m_\sigma$
in Table~\ref{tab:MMT}.
%


\end{document}